\documentclass[aps,prb,twocolumn,superscriptaddress,showpacs]{revtex4-1}

\usepackage{graphicx}
\usepackage{calc}
\usepackage{bm}
\usepackage{color}
\usepackage{mathtools}

\begin{document}

\title{Andreev reflection for MnTe altermagnet candidate}

\author{D.Yu.~Kazmin}
\author{V.D. Esin}
\author{Yu.S. Barash}
\author{A.V.~Timonina}
\author{N.N.~Kolesnikov}
\author{E.V.~Deviatov}
\affiliation{Institute of Solid State Physics of the Russian Academy of Sciences, Chernogolovka, Moscow District, 2 Academician Ossipyan str., 142432 Russia}

\date{\today}

\begin{abstract}
We experimentally study electron transport across a single planar junction between the indium electrode and MnTe altermagnet candidate. We confirm standard Ohmic behavior with strictly linear current-voltage curves above the indium critical field or temperature, although with high, about 100~kOhm, junction resistance. At low temperatures and in zero magnetic field,  we observe a well-developed Andreev curve with the pronounced coherence peaks, which cannot be normally expected for these high values of normal junction resistance. The conclusion on the Andreev reflection is also supported by suppression in magnetic field, as well as by universality of the observed behavior for all of the investigated samples. The experimental results can be explained by specifics of Andreev transport through the disordered region at the superconductor-altermagnet interface. Due to a different set of restrictions on the possibility of Andreev reflection,  an altermagnet suffers from the presence of disorder less than a normal spin-degenerate metal, so the conductance enhancement is retained throughout the superconducting gap. 
\end{abstract}

\maketitle

\section{Introduction}

Recently, a new class of altermagnetic materials has been added to  usual  ferro- and antiferro- magnetic classes~\cite{alter_common,alter_mazin}. Normally, ferromagnetic and antiferromagnetic materials  belong to the non-relativistic groups of magnetic symmetry, i.e. to the case of weak spin-orbit coupling. In contrast, topological materials are always characterized by strong spin-orbit interaction~\cite{Volkov-Pankratov,MZHasan,Armitage}, and, therefore, by spin-momentum locking~\cite{sm-valley-locking}. For example,  spin is rotating along the Fermi-arc in  Weyl semimetals~\cite{Fermi arc-SOC}, while the drumhead surface states lead to the spin textures of the skirmion type in topological nodal-line semimetals~\cite{nodal-line}. 

In altermagnets, the concept of spin-momentum locking was extended to the case of weak spin-orbit coupling~\cite{sm-valley-locking}, i.e. to the non-relativistic groups of magnetic symmetry~\cite{alter_common,alter_mazin}. As a result, the small net magnetization is accompanied by alternating spin-momentum locking in the k-space, so the unusual spin splitting is predicted~\cite{alter_common,alter_josephson}. For example, RuO$_2$ altermagnet consists of two spin sublattices with orthogonal spin directions~\cite{AHE_RuO2}. In the k-space, the up-polarized subband can be obtained by   $\pi/2$ rotation of the down-polarized subband, so RuO$_2$ altermagnet is characterized by d-wave order parameter~\cite{alter_supercond_notes,alter_normal_junction}. The  probability to scatter between subbands depends both on the electron spin and the propagation direction due to the spin-momentum locking~\cite{AHE_MnTe1}.   As a main experimental proposal, anomalous Hall effect~\cite{Armitage}  is predicted for altermagnets~\cite{alter_original}, despite of the weak net magnetization~\cite{alter_mazin,satoru}, as it has been experimentally demonstrated~\cite{AHE_RuO2,AHE_MnTe1,AHE_MnTe2,AHE_Mn5Si3} for the altermagnetic candidates MnTe, Mn$_5$Si$_3$, and RuO$_2$. 

In proximity to a superconductor, topological materials exhibit nontrivial physics due to the spin-momentum locking, that can in various cases result in topological superconductivity and existence of Majorana modes ~\cite{review1,review2,LiXu2019,volovik}. This concerns not only the topological insulators~\cite{zhang1,kane,zhang2,Fu}, but also Weyl semimetals, where the proximity
was predicted to produce specular Andreev reflection~\cite{spec}, similar to the graphene case~\cite{been1,been2}, as well as various superconducting pairings decaying in the depth of the sample~\cite{dutta2020}. 

Until now, superconductivity in altermagnets has only been studied   theoretically~\cite{alter_supercond_notes}. For example, superconductivity can  appear at high magnetic fields from a parent zero-field normal state of an altermagnet~\cite{AM_field_supercond}. Without external field, orientation-dependent effects are predicted for different superconductor-altermagnet-superconductor~\cite{alter_josephson,alter_josephson1} (SNS) and superconductor-altermagnet~\cite{SN1,SN2,SN3,alterSN} (SN) structures.  In a generic ferromagnet, as opposed to a traditional antiferromagnet, the number of conductivity channels for two spins are not the same, so the conventional Andreev conductivity~\cite{andreev,tinkham} is suppressed. Although, peculiarities of the Andreev reflection, related to the Fermi surface structure, can accur at interfaces between an itinerant antiferromagnet and an s-wave or d-wave superconductor~\cite{barash1,barash2,barash3,barash4}.  An altermagnet sometimes behaves as an antiferromagnet, and sometimes as a ferromagnet, depending on the interfacial orientation of the altermagnet relative to the superconductor. Therefore, possible qualitative modifications of the Andreev reflection can be expected in the case of altermagnets. 

 The disordered SN interface has also been considered for different altermagnet orientations~\cite{alterSN,SN1}. For an altermagnet,  the constant energy contours are ellipses with the major axes placed at some angle to the SN interface. Due to the spin-momentum locking and the Fermi wavevector mismatch, both Andreev and normal reflection depend on the SN interface orientation, so an altermagnet suffers from the presence of disorder less than a normal spin-degenerate metal~\cite{alterSN,SN1}.

Experimental investigations of a proximity-induced superconductivity can be conveniently performed for MnTe  altermagnetic candidate. The MnTe material has been well studied both experimentally~\cite{MnTe1,MnTe2,MnTe3,MnTe4,MnTe5,MnTe6,mntemag} and theoretically~\cite{MnTe_Mazin}. The  important point,  MnTe is characterized by relatively high conductance even at low temperatures~\cite{AHE_MnTe1,AHE_MnTe2} in contrast to most of  altermagnetic materials.

Here,  we experimentally study electron transport across a single planar junction between the indium electrode and MnTe altermagnet candidate. We confirm standard Ohmic behavior with strictly linear current-voltage curves above the indium critical field or temperature, although with high, about 100~kOhm, junction resistance. At low temperatures and in zero magnetic field,  we observe a well-developed Andreev curve with the pronounced coherence peaks, which cannot be normally expected for these high values of normal junction resistance. The conclusion on the Andreev reflection is also supported by suppression in magnetic field, as well as by universality of the observed behavior for all of the investigated samples. The experimental results can be explained by specifics of the  proximity-induced superconductivity in MnTe altermagnet.

\section{Samples and technique}

\begin{figure}
\includegraphics[width=1\columnwidth]{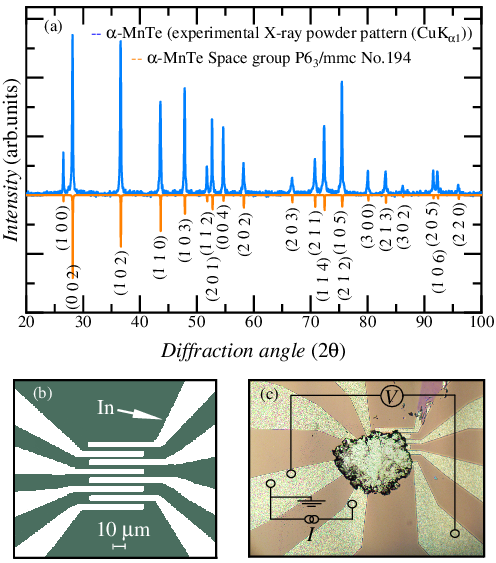}
\caption{(Color online) (a) The X-ray powder diffraction  pattern (Cu K$_{\alpha1}$ radiation), which is obtained for the crushed MnTe single crystal. The single-phase  $\alpha$-MnTe is confirmed with the space group $P6_3 /mmc$ No. 194.  (b) 5~$\mu$m width In leads  are formed by lift-off technique on a standard oxidized silicon substrate. (c) Image of the sample with electrical connections.  Single crystal MnTe flake is transferred to the In leads pattern, so planar In-MnTe junctions  are formed at the bottom surface of the flake. We study electron transport across a single In-MnTe junction in a standard three-point technique. 
  }
\label{fig1}
\end{figure}

MnTe was synthesized by reaction of elements (99.99\% Mn  and 99.9999\% Te) in evacuated silica ampules slowly heated up to 1050--1070$^\circ$C. The obtained loads were melted in the graphite crucibles under 10 MPa argon pressure, then homogenized at 1200$^\circ$C for 1 hour. The crystals grown by gradient freezing method are groups of single crystal domains with volume up to 0.5--1.0~cm$^3$. The MnTe composition is verified by energy-dispersive X-ray spectroscopy. The powder X-ray diffraction analysis confirms single-phase $\alpha$-MnTe with the space group $P6_3 /mmc$ No. 194, see Fig.~\ref{fig1} (a). 

The quality of our MnTe material was also tested in magnetization measurements~\cite{mntemag}.
MnTe  is an intrinsic room-temperature magnetic semiconductor with a collinear antiparallel magnetic ordering of Mn moments~\cite{MnTe1,MnTe2,MnTe3,MnTe4,MnTe5,MnTe6}.
The known for pure (stoichiometric) MnTe magnetic susceptibility drop~\cite{Xi_step,Xi_step1}  has been confirmed  around 80~K in zero magnetic field, as well as overall magnetization behavior~\cite{mntemag}.

Single-crystal samples are preferable in the fundamental research. For MnTe single crystal, there is no definite cleavage plane, so it is impossible to exfoliate several monolayers, similarly to standard graphene sample preparation technique. On the other hand, only three-dimensional MnTe is predicted as altermagnetic~\cite{MnTe_Mazin}.  Thus, we have to select  relatively thick (above 1~$\mu$m) MnTe single crystal flakes, which also ensures sample homogeneity.

Thick flakes requires special contact preparation technique~\cite{aunite,inwte1,inwte2,incosns,infgt,ingete}: the 100~nm In thick  leads are firstly formed by lift-off  on a standard oxidized silicon substrate to define the experimental geometry, as depicted in Fig.~\ref{fig1} (b). The parallel In stripes are of 5~$\mu$m width, they are separated by 2~$\mu$m intervals. As a second step, the fresh mechanically exfoliated MnTe flake is transferred to the In leads pattern. To produce In-MnTe junctions, the flake is shortly pressed  to the leads by another oxidized silicon substrate, the latter is removed afterward. As a result,   planar In-MnTe junctions  are formed at the bottom surface of the MnTe single crystal flake, with approximately $5\times 50$~$\mu$m$^2$ geometrical junction area for the best junctions, see Fig.~\ref{fig1} (c). This procedure usually provides transparent SN interfaces, stable in different cooling cycles, which has been verified  before for a wide range of materials~\cite{aunite,inwte1,inwte2,incosns,infgt,ingete}.  As an additional advantage, every In-MnTe interface  is protected from any contamination, since the junctiona are placed at the bottom side of a thick MnTe flake in Fig.~\ref{fig1} (c).  

We study electron transport across a single In-MnTe junction in a standard three-point technique: one In contact is grounded, the neighboring (2~$\mu$m separated) one is used as a voltage probe, while current is fed through another contact, as schematically presented in Fig.~\ref{fig1} (c). We use an additional (the fourth) wire to the grounded indium lead,  so all the wire resistances are excluded.  To obtain differential $dV/dI(I)$ characteristics, dc current $I$ is additionally modulated by a low (100~nA) ac component. We measure ac  voltage component ($\sim dV/dI$)  with a lock-in amplifier. The signal is confirmed to be independent of the modulation frequency below 100 Hz, which is defined by the applied filters. The dc current values are much smaller than the critical current for the indium leads, which can be estimated as  $\approx 30$~mA  for the leads' dimensions and the known~\cite{in-current} indium critical current density $j\approx 3\times 10^6$A/cm$^2$.

The indium leads are superconducting below the  critical temperature~\cite{indium} $T_c\approx 3.4~K$, so some of the measurements can be performed in a standard He$^4$ cryostat. Most of the results below are obtained within the 30~mK -- 1.2~K temperature range in a dilution refrigerator equipped with a superconducting solenoid.

 \begin{figure}
\includegraphics[width=\columnwidth]{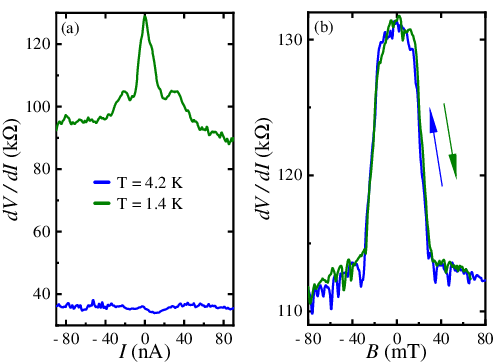}
\caption{(Color online) (a) Differential resistance $dV/dI$ of a single In-MnTe junction for 4.2~K and 1.4~K temperatures. At high 4.2~K temperature, $dV/dI$ is nearly independent of the applied current, demonstrating standard Ohmic behavior.  At low 1.4~K temperature, $dV/dI(I)$ curve is strongly nonlinear within $\pm40$~nA current range: differential resistance shows symmetric zero-bias peak with two satellite ones. (b) Magnetic field dependence of the zero-bias resistance peak, which is suppressed above $\pm 40$~mT normal to the interface magnetic field. The curves are shown for two sweep directions. Thus,  In-MnTe junctions show standard Ohmic behavior above the indium critical field or temperature, while below these values $dV/dI(I)$ curves strongly resemble Andreev reflection~\cite{andreev} at the disordered interface~\cite{BTK,tinkham}.
 }
\label{fig2}
\end{figure}

\section{Experimental results}

Fig.~\ref{fig2} (a) shows differential resistance $dV/dI$ of a single In-MnTe junction for 4.2~K and 1.4~K temperatures. At high 4.2~K temperature, $dV/dI$ is nearly independent of the applied current, demonstrating standard Ohmic behavior, see Fig.~\ref{fig2} (a). 

In the three-point technique, the measured resistance is the sum of the In-MnTe junction resistance, and the ones for the indium lead and for a some part of the bulk MnTe. If the first term dominates, the potential probe mostly reflects the voltage drop across a single In-MnTe interface, i.e. it reflects charge transfer between MnTe altermagnet and In superconductor. 

The measured resistance is about 40~kOhm in Fig.~\ref{fig2} (a), which is much higher than the bulk MnTe resistance.  The latter can be estimated as below 100~Ohm for our 5~$\mu$m  intervals from direct four-point measurements, which well corresponds to the known values~\cite{AHE_MnTe1,AHE_MnTe2}. Since the metallic indium is also of low resistance, the measured  $dV/dI$ only reflects the transport parameters of the In-MnTe interface. This conclusion is also verified by the  $dV/dI$ independence of the mutual positions of the current and voltage contacts. In contrast to the similar indium contacts to topological semimetals~\cite{aunite,inwte1,inwte2,incosns,infgt,ingete}, the measured resistance is quite high for the $5\times 50$~$\mu$m$^2$ junction area, probably due to the band bending~\cite{shklovskii} and/or disorder~\cite{batov} at the interface.  

At low 1.4~K temperature, indium leads are in a superconducting state~\cite{indium}.  The In-MnTe junction resistance is even increased to  about 100~kOhm value, see Fig.~\ref{fig2} (a). However, the experimental  $dV/dI(I)$ curve is strongly nonlinear now: differential resistance is symmetrically increased within $\pm 40$~nA, resembling standard Andreev reflection~\cite{andreev,tinkham} at the disordered interface~\cite{BTK,nbsemi,ingasb}. This analogy is also confirmed by the magnetic field dependence: the enhanced zero-bias resistance only survive within $\pm 40$~mT magnetic field, see Fig.~\ref{fig2} (b), which is close to the indium critical field~\cite{indium}. Thus, we wish to emphasize that the  $dV/dI(I)$ non-linearity only appears for temperatures and magnetic fields below  the superconducting indium transition in Fig.~\ref{fig2}. 

Usually, one cannot expect Andreev reflection for strongly resistive interface: even if we considered our planar In-MnTe junction as a one-channel point contact with $h/e^2$ channel resistance, it would be much smaller than the obtained 90~kOhm value to both sides of the supposed superconducting gap. Also, for the 90~kOhm junction, the $0.5$~mV indium superconducting gap~\cite{indium} corresponds to the $\pm 5.5$~nA current range in Fig.~\ref{fig2} (a), while the experimental $\pm 40$~nA value is one order of magnitude higher.

\begin{figure}
\includegraphics[width=\columnwidth]{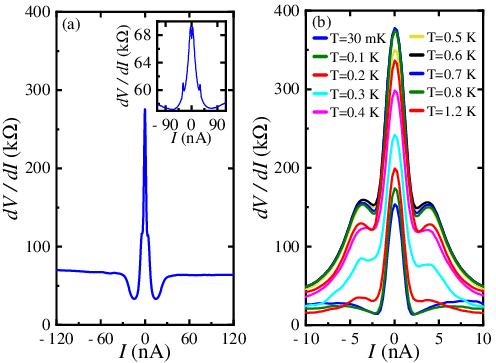}
\caption{(Color online)  (a) $dV/dI(I)$ curve in a wide current range at 1.2~K temperature for In-MnTe junction with 75~kOhm normal resistance $R_N$. There is sharp $dV/dI$  peak at zero bias, which is accompanied by two symmetrical $dV/dI$ minima (also known as coherence peaks in differential conductance). At higher current values, differential resistance is stable at $\approx$75~kOhm in a wide current range. This behavior is exactly that one expect for the disordered Andreev interface~\cite{BTK,tinkham}. The inset shows qualitatively similar behavior for another, more transparent In-MnTe junction with about 60~kOhm normal resistance. Both the junctions show two additional subgap peaks at low biases, similarly to one in Fig.~\ref{fig2}.  (b) Temperature dependence of $dV/dI(I)$ curves for the $R_N\approx75$~kOhm sample. This temperature dependence is non-monotonic, the zero-bias peak is increasing from 1.2~K to 0.7~K, afterward it is sharply decreasing for temperatures below 0.6~K. The positions of the subgap  peaks are crudely independent of temperature, while their relative magnitude seems to be decreasing. The data are obtained in zero magnetic field. 
 }
\label{fig3}
\end{figure}  

Despite these contradictions, the conclusion on the Andreev reflection at the In-MnTe interface is also supported by universality of the observed behavior, see Fig.~\ref{fig3} (a) for two different samples at 1.2~K. 

In the main field of Fig.~\ref{fig3} (a), there is sharp $dV/dI$  peak at zero bias, which is accompanied by two symmetrical $dV/dI$ minima (also known as coherence peaks in differential conductance~\cite{tinkham}). At higher current values, differential resistance is stable at $\approx$75~kOhm in a wide current range.  This symmetric behavior is exactly that one expect for the disordered Andreev interface~\cite{BTK,tinkham}, while it is inconsistent with single-particle potentials, e.g. with Schottky barrier at the surface of a semiconductor. The inset to  Fig.~\ref{fig3} (a) shows qualitatively similar behavior for more transparent In-MnTe junction: the normal junction resistance is about 60~kOhm (still above $h/e^2$), there are two symmetrical resistance minima (conductance coherence peaks), while the zero-bias peak resistance is much smaller than in the main field of  Fig.~\ref{fig3} (a).  Similarly to Fig.~\ref{fig2} (a), the coherence peaks' positions correspond to the  4.5~mV and 1.1~mV voltage bias for the inset and the main field of  Fig.~\ref{fig3} (a), respectively, these values are also too high for the 0.5~mV indium superconducting gap. 

All the junctions show several additional subgap conductance peaks, i.e. at biases smaller than the positions of the main coherence peaks, see Fig.~\ref{fig2} (a), the main field of Fig.~\ref{fig3} (the subgap peaks are better seen in the (b) panel due to the smaller current range), and the inset to Fig.~\ref{fig3}. The peaks can not be connected with multiple Andreev reflection~\cite{tinkham,BTK}, since it requires a short $L<\xi$ highly transparent SNS contact, where $\xi$ is a coherence length.  The positions of the subgap  peaks are crudely independent of temperature in Fig.~\ref{fig3} (b), while their relative magnitude seems to be decreasing for lower temperatures. 

\begin{figure}
\includegraphics[width=\columnwidth]{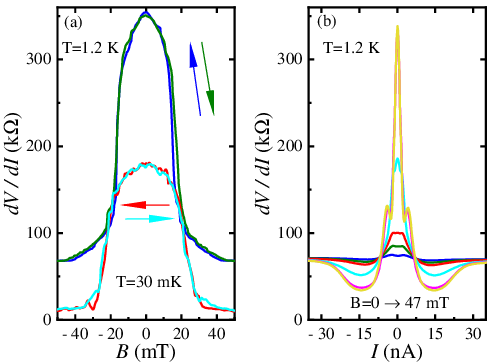}
\caption{(Color online) Magnetic field dependence of differential resistance for the $R_N\approx75$~kOhm sample from Fig.~\ref{fig3}.  (a) The zero bias resistance  is suppressed by the $\pm 40$~mT normal to the interface magnetic field. The data are shown for two temperatures, 1.2~K and 30~mK, respectively, and for two sweep directions of the magnetic field, as indicated by arrows. The curves are symmetric in magnetic field, there is no hysteresis with the field sweep direction. (b) $dV/dI(I)$ curves for different magnetic fields at 1.2~K. The field values are 0, 8~mT, 15~mT, 22~mT, 30~mT, and 47~mT. The $dV/dI(I)$ curve nonlinearity is monotonically suppressed by magnetic field to the flat curve at  $\pm 47$~mT. The standard Ohmic behavior is recovered above the indium critical field even at low temperature, so the magnetic field dependence supports Andreev origin of nonlinear  $dV/dI(I)$ curves.
  }
\label{fig4}
\end{figure}

Usually, one can not expect temperature dependence much below the indium superconducting gap. In contrast, Fig.~\ref{fig3} (b) shows nonmonotonic $dV/dI(I)$ behavior below 1~K: for low biases, differential resistance is increasing from 1.2~K to 0.7~K, afterward it is sharply decreasing for temperatures below 0.6~K. The effect of temperature can be also seen from the $dV/dI(B)$ magnetic field  scans at two different temperatures in  Fig.~\ref{fig4} (a). Both the subgap resistance and the normal one are diminishing from 1.2~K to 30~mK temperature. 

Despite the mentioned $dV/dI(I)$ peculiarities (high junction resistance, incorrect gap value  and the subgap conductance peaks), the $dV/dI(I)$ non-linearity is induced by superconductivity, as it is strongly confirmed by magnetic field dependence in Fig.~\ref{fig4}. The zero bias resistance  is suppressed by the $\pm 40$~mT magnetic field, see  Fig.~\ref{fig4} (a), the  $dV/dI(I)$ curve is flat above  $\pm 40$~mT magnetic field, see  Fig.~\ref{fig4} (b), so the standard Ohmic behavior is recovered above the indium critical field even at low temperature.

\section{Discussion} \label{disc}

At low temperatures and in zero magnetic field,  we observe a well-developed Andreev curve with the pronounced coherence peaks, which cannot be normally expected for these high values of normal junction resistance.

First of all, one can be sure that $R_N\sim$100~kOhm is due to the In-MnTe interface, since the bulk MnTe resistance is below 100~Ohm for our 5~$\mu$m intervals~\cite{AHE_MnTe1,AHE_MnTe2}, as confirmed by direct resistance measurements. The quality of our MnTe material was also tested in magnetization measurements~\cite{mntemag}, the bulk MnTe composition was verified by energy-dispersive X-ray spectroscopy and the powder X-ray diffraction analysis, see above. Universality of the normal resistance  for three different samples also  indicates that $R_N\sim$100~kOhm should not  be connected with  any bulk MnTe disadvantages~\cite{impur1,impur2}. 

On the other hand, MnTe  is an intrinsic room-temperature magnetic semiconductor with finite band gap~\cite{MnTe1,MnTe2,MnTe3,MnTe4,MnTe5,MnTe6}. Because of the edge  electrostatics~\cite{shklovskii}, the carriers' concentration varies smoothly by approaching the crystal surface, so there is a depletion region of finite width at the In-MnTe interface~\cite{shklovskii,kouwenhoven}. For a three-dimensional case, this depletion region is too narrow to distort Ohmic behavior of the normal junction resistance (i.e. above indium critical field or temperature), so we observe linear Ohmic $dV/dI(I)$ curves in Figs.~\ref{fig2} (a)  and~\ref{fig4} (b). However, being placed between normal and superconducting electrodes, the depletion region strongly affects the transport properties of SN junction, as it was experimentally demonstrated, e.g. in  Ref.~\onlinecite{batov}. The coherence peaks' positions are thus strongly affected by the voltage drop over this depletion region, leading to the incorrect gap determination from the $dV/dI(I)$ curves, as we obtain in Figs.~\ref{fig2} and~\ref{fig3}.

For the extended disordered region, it is demonstrated the presence of the subgap resonant states through combination of normal and Andreev reflection processes~\cite{alterSN}.  The presence of satellite subgap peaks in Figs.~\ref{fig2} and~\ref{fig3} can thus be explained by these resonance states~\cite{alterSN}. The conclusion on the disorder-induced effects is also confirmed by temperature dependence in Fig.~\ref{fig3} (b). While the 40~mT magnetic field suppress superconductivity for indium electrodes in in Figs.~\ref{fig2} (b) and~\ref{fig4}, one can not expect any temperature dependence neither for indium nor for the MnTe spectrum well below 1~K. Instead, the impact of disorder is diminishing for low temperatures due to the higher mean free path, so the In-MnTe junction conductance is increasing both outside and within the superconducting gap. 

As a main experimental result, we observe a well-developed Andreev curve with the pronounced coherence peaks for high values of normal junction resistance $R_N\sim$100~kOhm.  Even if we considered our planar In-MnTe junction as a one-channel point contact with $R_N=T \times h/e^2$ channel resistance, $R_N\sim$100~kOhm value  would correspond to the $T=h/e^2/R_N \sim 0.25$ Landauer transmission probability of this channel. In the terms of BTK theory~\cite{BTK}, it corresponds to the tunnel In-MnTe junction with the BTK~\cite{BTK} barrier strength $Z=(1/T-1)^{1/2}\approx 2$, which does not correlate with the finite subgap resistance $R_s$. The real transmission  should be much smaller for the planar $5\times 50$~$\mu$m$^2$ In-MnTe junctions with multiple parallel channels, so the contradiction is even stronger between a well-developed Andreev curve and high values of normal junction resistance.

Irrespective to the number of conductive channels, Andreev reflection is a two-particle process~\cite{andreev,tinkham}. Thus, if the normal conductance $1/R_N$ is defined by the transmission probability $T$, the  subgap resistance $R_s$ should be proportional to $T^{-2}$, so the single-particle transmission $T$ can be estimated as $T\sim R_N/R_s$ for $T<1$. While the normal resistance $R_N$ is quite universal for different junctions, difference in $R_s$ gives  $T\approx$0.7 in Fig.~\ref{fig2} ($Z\approx 0.7$),  0.85 in the inset to Fig.~\ref{fig3} (a) ($Z\approx 0.4$), and 0.2 in  Figs.~\ref{fig3} and~\ref{fig4} ($Z\approx 2$). This simple estimation can be confirmed by a direct fitting of the experimental $dV/dI(I)$ curves by the BTK equation~\cite{BTK}. It gives  $Z=0.72$ in Fig.~\ref{fig2},  0.65 in the inset to Fig.~\ref{fig3} (a), and 1.32 in  Figs.~\ref{fig3} and~\ref{fig4}. These low $Z$ values does not correlate with ones, obtained from the normal junction resistance $R_N$, even if one considered $5\times 50$~$\mu$m$^2$ planar junction as one-channel point contact. 

Thus, one can not expect to observe finite subgap resistance for planar junctions with high $R_N\sim$100~kOhm, irrespective of model BTK delta-potential~\cite{BTK}   or real extended disordered depletion region at the In-MnTe interface.

MnTe belongs to a new class of altermagnetic materials~\cite{MnTe_Mazin,AHE_MnTe1,AHE_MnTe2}, the small net magnetization is accompanied by alternating spin-momentum locking in the k-space, so the unusual spin splitting is predicted~\cite{alter_common,alter_josephson}. The antiferromagnetic ordering with vanishing net magnetization is known for $\alpha$-MnTe below 307--325~K, depending on the thin films or the single crystal samples~\cite{NeelTemp_SC,Xi_step1,MnTe1}.  The reorientation field of the N\'eel vector was found to be between 2 and 3~T for MnTe~\cite{AHE_MnTe1,AHE_MnTe2}.   Thus, we should consider bulk MnTe as altermagnetic in its ground state for our temperature and magnetic field ranges.

The disordered SN interface has been considered in Refs.~\cite{alterSN,SN1} for different altermagnet orientations, the behavior of the altermagnetic materials in this case is different than the one of normal spin-degenerate metal. In the latter case, the effect of disorder can be compared to the effect of a tunneling barrier, with a strong resonant peak at the gap edge and quick decay of conductance within the gap. For an altermagnet,  the constant energy contours are ellipses with the major axes placed at some angle to the SN interface. Due to the spin-momentum locking, the Fermi wavevector mismatch varies for two different spin channels, so even for a delta function barrier at the interface, altermagnet retains the conductance enhancement throughout the superconducting gap~\cite{alterSN,SN1}. In other words, altermagnet suffers less from the presence of disorder in comparison with normal spin-degenerate metal due  to a different set of restrictions on the possibility of Andreev reflection~\cite{alterSN,SN1}. 

It is obvious, that the restrictions on the possibility of Andreev reflection depend on the orientation of the spin-splitting direction.  In all of the cases however, the behavior outside of the superconducting gap is largely equivalent~\cite{alterSN}. In our experiment, since for MnTe there is no definite cleavage plane, the MnTe single crystal orientation is arbitrary in respect to the In plane. However, all the samples show similar values of normal junction resistance $R_N$, as it can be expected for an altermagnet~\cite{alterSN}.  In contrast,  the single-particle transmission $T$ varies from 0.2 to 0.85, as obtained from the $R_s/R_N$ ratio, so this difference can be attributed to orientational  dependence of Andreev reflection for altermagnets~\cite{SN1,SN2,SN3,alterSN}. 

It is worth mentioning, that recent theoretical results predict that proximity effects in two-dimensional altermagnet/s-wave superconductor junctions induce a singlet-triplet superconducting order parameter, featuring a d-wave-like singlet component~\cite{dwaveS}. The physics of such junctions is sensitive to the relative strength of the d-wave-like magnetic term in the altermagnet compared to the s-wave order parameter of the adjacent superconductor. Possible effects include significant conductance anisotropy and the appearance of a mirage gap~\cite{mirage_gap}. Although these results are not directly applicable to our case due to their destruction by disorder and sensitivity to the dimensionality of the device, they highlight potential complications of the analysis of the transport properties of the system. Similar theory for three-dimensional altermagnet/s-wave superconductor junctions, including the disordered interface region, is still lacking.

\section{Conclusion}
As a conclusion, we experimentally study electron transport across a single planar junction between indium electrode and MnTe altermagnet candidate. We confirm standard Ohmic behavior with strictly linear current-voltage curves above the indium critical field or temperature, although with high, about 100~kOhm, junction resistance. At low temperatures and in zero magnetic field,  we observe a well-developed Andreev curve with the pronounced coherence peaks, which cannot be normally expected for these high values of normal junction resistance. The conclusion on the Andreev reflection is also supported by suppression in magnetic field, as well as by universality of the observed behavior for all of the investigated samples. The experimental results can be explained by specifics of Andreev transport through the disordered region at the superconductor-altermagnet interface.

\acknowledgments

We wish to thank S.S~Khasanov for X-ray sample characterization and Vladimir Zyuzin for valuable discussions.  We gratefully acknowledge financial support  by the  Russian Science Foundation, project RSF-24-22-00060, https://rscf.ru/project/24-22-00060/

\end{document}